\documentstyle[prb,eqsecnum,aps]{revtex}
\begin{document}
\title{Looking at the Haldane Conjecture from a Grouptheoretical Point of View}

\author{K.-H. M\"utter\cite{muetter}}
\address{Physics Department, University of Wuppertal\\
42097 Wuppertal, Germany}

\date{01.03.94}

\maketitle
\begin{abstract}
Based on the Lieb-Schultz-Mattis construction we present a five parameter
family of Spin-1 Hamiltonians with degenerate groundstate. Starting from the
critical $SU(3)$ symmetric Hamiltonian, we look for those perturbations of the
$SU(3)$ symmetry, which leave the groundstate degenerate. We also discuss the
spin-3/2 $SU(4)$-case.

\end{abstract}
\draft
\section{Introduction}
The critical properties of the spin-$1/2$ XYZ model with Hamiltonian:
\begin{equation}
	H(J_1,J_2,J_3)=\sum_{l=1}^3 J_lH_l,\quad
	H_l=\sum_{x=1}^N \sigma_l(x)\sigma_l(x+1)
\end{equation}
are wellknown. The $SU(2)$ symmetric model with isotropic couplings
$J_1=J_2=J_3$ is critical in the sense, that there is no gap between
the groundstate and the lowlying excited states. \cite{lsm} The spectrum of
lowlying energy eigenvalues changes, if we perturb the $SU(2)$ symmetry. One
has to distinguish two types of perturbations:
\begin{itemize}
\item[(1)]
 	longitudinal perturbations:
	\begin{equation}
		J_1=J_2=J,\quad J_3=J+\Delta J_3.
	\end{equation}
	The symmetries of this one parameter family of Hamiltonians allow for
	the construction \cite{lsm} of a unitary operator $U$:
	\begin{equation}
		U=\exp(iA),\quad
		A=\frac{\pi }{N}\sum_{x=1}^Nx\sigma_3(x),
	\end{equation}
	which creates from the groundstate $|0\rangle$ a new orthogonal state
	\begin{equation}
		|1\rangle=U|0\rangle,\quad \langle0|1\rangle =
		\langle0|U|0\rangle=0,
	\end{equation}
	with the same energy in the thermodynamical limit
	$N\rightarrow \infty$
	\begin{equation}
		\langle1|H(J,J,J_3)|1\rangle-\langle0|H(J,J,J_3)|0\rangle =
		O(N^{-1}).
	\end{equation}
	Therefore the groundstate is at least twofold degenerate. Indeed, the
	system with Hamiltonian (1.2) remains critical for perturbations
	$\Delta J_3 < 0$ [Ref. \onlinecite{yy}]. For $\Delta J_3 > 0$, there
	is a twofold degenerate groundstate and a gap between the lowest
	energy eigenvalues in the sectors with total spin $|S_3|=1$ and
	$S_3=0$:
	\begin{equation}
		E(|S_3|=1,\Delta J_3)-E(|S_3|=0,\Delta J_3)=
		16\pi\exp\left(-\frac{\pi^2}{\sqrt{8\Delta J_3}}\right).
	\end{equation}
	The gap vanishes exponentially in the isotropic limit
	$\Delta J_3\rightarrow 0$ [Ref. \onlinecite{cloi}]. Therefore at the
	$SU(2)$  symmetric point, where the Hamiltonian (1.1) can be expressed
	in terms of permutation operators:
	\begin{equation}
		P(x,y)=\frac{1}{2}\openone +
		\frac{1}{2}\vec{\sigma}(x)\vec{\sigma}(y),
	\end{equation}
	there is a phase transition of the Kosterlitz-Thouless type in the
	anisotropy parameter $\Delta J_3$.
\item[(2)]
 	transverse perturbations:
	\begin{equation}
		J_1=J+\Delta J,\quad J_2=J-\Delta J,\quad J_3=J.
	\end{equation}
	The groundstate is now unique and a gap opens between the groundstate
	and the first excited state.
\end{itemize}

It is the purpose of this paper to investigate the critical properties
of higher spin models with $s=1 $ and $ s=\frac{3}{2}$. The first
question we should answer is:
``What is the 'natural' extension of the $XYZ$-model (1.1) with
broken $SU(2)$ symmetry in case of higher spins ?''

Usually one substitutes the spin-$1/2$ matrices by the corresponding
spin-$1$,-$3/2$ matrices. Haldane's conjecture \cite{haldane} tells us that
in this naive extension we compare Hamiltonians with completely different
critical properties: In contrast to the isotropic spin-$1/2$ model, the
isotropic spin-$1$ model is predicted to have a unique groundstate and a gap.
For reasons, which will become clear in this paper, the natural extensions
of the $XYZ$-model are given by Hamiltonians which we construct from
the generators $\lambda_A,A=1,2,...,n=m^2-1$ of the groups $SU(m),m=3,4,...$ :
\begin{equation}
	H(J_A,A=1,...,n)=\sum_{A=1}^n J_AH_A,\quad
	H_A=\sum_{x=1}^N \lambda_A(x)\lambda_A(x+1).
\end{equation}
The number of anisotropy parameters $J_A$ is given by the number
$n=m^2-1$ of generators in the internal symmetry group $SU(m)$.
This means $n=8$ for
the spin-$1$ case and $n=15$ for the spin-$3/2$ case. Looking for the
critical properties in these high dimensional parameter spaces, we got
the impression, that the various spin cases $s=1/2,1,3/2$ are not that
different.

It is the main goal of this paper to learn as much as possible on the
'critical submanifold' in the space of anisotropy parameters $J_A$, where
the systems with Hamiltonians (1.9) are gapless. The isotropic point:
\begin{equation}
	J_A=J,\quad A=1,...,n=m^2-1,
\end{equation}
is known to lie on the critical submanifold for all internal symmetry
groups $SU(m)$ [Ref. \onlinecite{sutherland}]. It is also remarkable to note
that these Hamiltonians can always be expressed in terms of the permutation
operators:
\begin{equation}
 	P(x,y)=\frac{1}{m}\openone +
		\frac{1}{2}\sum_{A=1}^{m^2-1} \lambda_A(x) \lambda_A(y).
\end{equation}
Having found one point on the critical submanifold we are lead to our
second question:``What kind of perturbations of the critical $SU(m)$
symmetric Hamiltonian do not destroy criticality?''
We want to show in this paper that a partial answer can be found by a
generalization of the Lieb-Schultz-Mattis construction \cite{afflieb} for a
submanifold of Hamiltonians of the type (1.9).

The outline of the paper is as follows: In Sec. II we present the
Lieb-Schultz-Mattis construction for the spin-$1$ $SU(3)$-case . In
Sec. III we compare our results on the criticality of spin-$1$
models with the known results on the bilinear and biquadratic
spin-$1$ Hamiltonian. In Sec. IV we extend the Lieb-Schultz-Mattis
construction to the spin-$3/2$ $SU(4)$-case.
\section{The spin-1 $SU(3)$-case}
In this section, we are concerned with the Hamiltonians (1.9) for
the $SU(3)$-case. It is convenient to represent the $SU(3)$ generators
by the Gell-Mann matrices listed in (A.1). We are going to prove
that the Lieb-Schultz-Mattis construction \cite{afflieb} is possible for those
Hamiltonians (1.9) with couplings:
\begin{equation}
	J_1,J_2,J_3,J_4=J_5,J_6=J_7,J_8.
\end{equation}
The unitary operator $U$, which creates from the groundstate $|0\rangle$
a second orthogonal groundstate $|1\rangle=U|0\rangle$, is found to be:
\begin{equation}
	U=\exp(iA),\quad A=\frac{2\pi}{\sqrt{3}N}\sum_{x=1}^N x\lambda_8(x).
\end{equation}
The following symmetries of the Hamiltonian are needed:
\begin{itemize}
\item[(1)]
	 Reflection-and translation invariance R and T. Note that these
	two operators anticommute:
	\begin{equation}
		RT+TR=0.
	\end{equation}
	Therefore, momentum $p$ is a good quantum number and the groundstate
	is trivially degenerate due to reflection invariance if $p\neq 0,\pi$.
\item[(2)]
	Conservation of:
	\begin{equation}
		\Lambda_8=\sum_{x=1}^N\lambda_8(x).
	\end{equation}
	This property follows from the commutation relations with the various
	pieces of the Hamiltonian (2.1):
	\begin{equation}
		[\Lambda_8,H_A]=0, \quad A=1,2,3,8,
	\end{equation}
	\begin{equation}
		[\Lambda_8,H_4+H_5]=[\Lambda_8,H_6+H_7]=0.
	\end{equation}
	They can be easily derived from the properties of the Gell-Mann
	matrices (A.1). Note that the groundstate is an eigenfunction of:
	\begin{equation}
		\frac{1}{\sqrt{3}N}\Lambda_8|0\rangle =
		\frac{N-3N_0}{3N}|0\rangle.
	\end{equation}
	The eigenvalue counts the number $N_0$ of sites with ``spin 0''. The
	spin at each site $x$ is measured by the eigenvalue of
	$\lambda_3(x)$.
\end{itemize}

We are now in the position to show that the difference of the expectation
values:
\begin{equation}
	\langle1|H|1\rangle-\langle0|H|0\rangle=O(N^{-1}),
\end{equation}
vanishes for the Hamiltonians (2.1) in the thermodynamical limit. Due to
the commutation relations (A.6) and the transformation properties
(A.9), (A.10) we get for the various pieces of the Hamiltonian (2.1)
\begin{equation}
	\langle0|UH_AU^+|0\rangle-\langle0|H_A|0\rangle=0, \quad A=1,2,3,8,
\end{equation}
\begin{equation}
	\langle0|U(H_4+H_5)U^+|0\rangle-\langle0|H_4+H_5|0\rangle=O(N^{-1}),
\end{equation}
\begin{equation}
	\langle0|U(H_6+H_7)U^+|0\rangle-\langle0|H_6+H_7|0\rangle=O(N^{-1}).
\end{equation}
In order to see the orthogonality of the two states $|0\rangle$ and
$|1\rangle=U|0\rangle$ we use translation invariance, the conservation of
$\Lambda_8$, the periodic boundary condition:
\begin{equation}
	\lambda_8(N+1)=\lambda_8(1),
\end{equation}
and the explicit form (A.1) of $\lambda_8(1)$
\begin{equation}
	\exp\left(\frac{2\pi i}{\sqrt{3}}\lambda_8(1)\right) =
	 \exp\left(\frac{2\pi i}{3}\right)\openone,
\end{equation}
\begin{equation}
	\langle0|U|0\rangle=\exp\left(2\pi i\frac{N_0}{N}\right)
	\langle0|U|0\rangle.
\end{equation}
Therefore, the two states are orthogonal provided that:
\begin{equation}
	\exp\left(2\pi i\frac{N_0}{N}\right)\neq 1.
\end{equation}
This means that in the groundstate $|0\rangle$ the fraction
$N_0/N $ of sites with spin $0$ is neither zero nor one.
We expect that this fraction is $1/3$ --at least in the $SU(3)$ symmetric case,
where the groundstate is built up from  direct products of three
fundamental representations:
\begin{equation}
	3 \otimes 3 \otimes 3 =1 \oplus 8 \oplus 8 \oplus 10.
\end{equation}
The singlet part is projected out by antisymmetrization. These ``trimer''
or ``baryon'' states are the $SU(3)$ analogue of the ``dimer'' or
``valence bond'' states in the $SU(2)$ case.\\
\section{The bilinear and biquadratic spin-1 Hamiltonians}
The spin-$1$ Hamiltonian:\cite{aff}
\begin{equation}
	H(\omega)=J\sum_x\left[\cos \omega\, \vec{s}(x)\vec{s}(x+1)+
	\sin \omega \biglb( \vec{s}(x)\vec{s}(x+1)\bigrb) ^2\right],
\end{equation}
built up from the $O(3)$ generators $s_l,l=1,2,3$ with bilinear and
biquadratic couplings is considered here as a special case of (1.9)
with a specific perturbation of the $SU(3)$-symmetry. From the
explicit form (A.1) of the Gell-Mann matrices we can identify:
\begin{equation}
	s_1(x)=\lambda_7(x),
	\quad s_2(x)=-\lambda_5(x),
	\quad s_3(x)=\lambda_2(x).
\end{equation}
Using the commutation and anticommutation relations (A.2) and (A.4)
we can express the biquadratic form
\begin{equation}\nonumber
	\biglb(\vec{s}(x)\vec{s}(x+1)\bigrb)^2
\end{equation}
in terms of bilinears $\lambda_A(x)\lambda_A(x+1)$ and arrive at the
following representation of the Hamiltonian (3.1):
\begin{equation}
	H(\omega)=J_1\sum_{A\neq 2,5,7}H_A+J_2\sum_{A=2,5,7} H_A,
\end{equation}
where
\begin{equation}
	J_2=J\left(\cos \omega-\frac{1}{2}\sin \omega\right),
	\quad J_1=\frac{J}{2}\sin \omega.
\end{equation}
The $SU(3)$ symmetric point is found at:
\begin{equation}
	\omega=\frac{\pi}{4},\quad J_1=J_2=J\frac{1}{2\sqrt{2}}.
\end{equation}
This is the only Hamiltonian (3.1) which belongs to the submanifold (2.1)
of Hamiltonians with degengerate groundstate. The Hamiltonian with pure
biquadratic coupling:
\begin{equation}
	\omega=\frac{\pi}{2},\quad J_2=-\frac{J}{2},\quad J_1=\frac{J}{2}
\end{equation}
does not belong to the  submanifold (2.1). The Hamiltonian (3.7) has
been proven by A.Kl\"umper \cite{klu} to have a gap.
The Hamiltonian (3.1) with:
\begin{equation}
	\omega=-\frac{\pi}{4}:\quad J_2=\frac{3}{2\sqrt{2}},\quad
                            J_1=-\frac{1}{2\sqrt{2}}
\end{equation}
has been proven to be critical.\cite{kulish} However it does not belong to the
submanifold (2.1). We therefore suggest, that there are
further unitary operators of the Lieb-Schultz-Mattis type. However,
so far we did not succeed to construct them.

We would like to stress here, that the Hamiltonians (3.5) and (3.8)
($\omega=\pm \pi/4$) are the only ones which have been proven
to be critical. Moreover, it has been suggested by I. Affleck, \cite{aff}
that they are indeed the only ones. E.g. the pure bilinear
Hamiltonian with $\omega=0$ has a gap and a unique groundstate,
if Haldanes conjecture is correct. From our group theoretical point
of view, the perturbation of the $SU(3)$-symmetry in the one
parameter family (3.1) is comparable with the transverse perturbations
(1.8) of the $SU(2)$-symmetry in the spin-$1/2$ case which destroys
criticality.
\section{The spin-3/2 $SU(4)$-case }
We are now considering Hamiltonians of the type (1.9) which are built up
from the $15$ generators $\lambda_A,A=1,...,15$ of the group $SU(4)$.
We are looking for the submanifold of Hamiltonians with degenerate
groundstate. The Lieb-Schultz-Mattis construction can be performed
for Hamiltonians with couplings
\begin{equation}
	J_A,\quad A=1,...,8,15 \quad J_9=J_{10},J_{11}=J_{12},J_{13}=J_{14}.
\end{equation}
The unitary operator $U$, which creates from the groundstate $|0\rangle$
a second orthogonal groundstate $|1\rangle=U|0\rangle$, is found to be:
\begin{equation}
	U=\exp(iA),\quad A =
	\frac{\pi\sqrt{3}}{N\sqrt{2}}\sum_{x=1}^N x\lambda_{15}(x).
\end{equation}
The proof is completely analogous to the $SU(3)$-case in Sec. II and is based
on reflection and translation invariance and the conservation of:
\begin{equation}
	\Lambda_{15}=\sum_{x=1}^N\lambda_{15}(x).
\end{equation}
This property follows from the commutation relations with the various
pieces of the Hamiltonian (4.1):
\begin{equation}
	[\Lambda_{15},H_A]=0, \quad A=1,...,8,15,
\end{equation}
\begin{equation}
	[\Lambda_{15},H_9+H_{10}]=[\Lambda_{15},H_{11}+H_{12}]
	= [\Lambda_{15},H_{13}+H_{14}]=0.
\end{equation}
Note that the groundstate is an eigenfunction of
$\Lambda_{15}$ and we expect its eigenvalue to be $0$.

In contrast to the $SU(3)$-case, we find within the submanifold (4.1)
a smaller submanifold of Hamiltonians with couplings
\begin{equation}
	J_1=J_2,J_3,J_4=J_5,J_6=J_7,J_8,J_9=J_{10},J_{11}=J_{12},J_{13}=J_{14},
	J_{15},
\end{equation}
where a second unitary operator
\begin{equation}
	W=\exp (iB),\quad B=\frac{2\pi }{N}\sum_{x=1}^N x s_3(x)
\end{equation}
creates from the groundstate $|0\rangle$ a new orthogonal state
$|2\rangle=W|0\rangle$ with the same energy. The operator $B$ is constructed
from the spin-$3/2$ matrices $s_3(x)$ which can be represented by a linear
combination of the commuting matrices $\lambda_3,\lambda_8,\lambda_{15}$.
The three states $|0\rangle,|1\rangle=U|0\rangle,|2\rangle=W|0\rangle$ with
the groundstate energy in the thermodynamical limit turn out to be orthogonal
to each other:
\begin{equation}
	\langle0|U|0\rangle=\langle0|W|0\rangle=\langle0|U^+W|0\rangle=0.
\end{equation}
The proof relies on the explicit form of $\lambda_{15}$ and $s_3$ which
yields:
\begin{equation}
	\exp\left(\pi i \sqrt{\frac{3}{2}} \lambda_{15}\right)=i
	\openone,\quad \exp(2\pi is_3)=-\openone.
\end{equation}
Thus we end up with the conclusion: The groundstates of the Hamiltonians
(4.6) are at least threefold degenerate!
\section{Conclusion}
In this paper we have made an attempt to attack the question:
``What are the crucial characteristics of critical (i.e. gapless) quantum
spin Hamiltonians with nearest neighbour couplings in one dimension?''
Though we are far from a complete answer, we would like to emphasize
on the following points:
\begin{itemize}
\item[(1)]
	$SU(m)$-invariant Hamiltonians \cite{sutherland} are critical.
	They can be expressed in terms of nearest neighbour permutation
	operators.
\item[(2)]
	Specific perturbations of the $SU(m)$ symmetry do not destroy
	criticality completely in the sense, that the groundstate remains
	degenerate at least to a certain degree (twofold, threefold, etc. ).
	For $m=2$ --i.e. $s=1/2$-- these perturbations are known to form a one
	dimensional submanifold -- the familiar spin-$1/2$ XXZ-models --
	in the two dimensional manifold of anisotropy parameters.
\item[(3)]
	For $m=3$ --i.e. $s=1$-- we found in the seven dimensional manifold
	of anisotropy parameters a five dimensional submanifold, where the
	groundstate is at least twofold degenerate.
\item[(4)]
	A special example of a one parameter family of spin-$1$ Hamiltonians
	with degenerate	groundstate is:
	\begin{equation}
		H(\Delta J_3)=\sum_x\left[\vec{s}(x)\vec{s}(x+1)+
 		\biglb( \vec{s}(x)\vec{s}(x+1) \bigrb)^2 +
		\Delta J_3 s_3(x)s_3(x+1)\right].
	\end{equation}
	Note that for the Hamiltonian (5.1)
	the numbers $N_0,N_1,N_{-1}$ of sites with spin $0,1,-1$ are conserved.
	In our opinion, the Hamiltonian (5.1) is the ``true''
	spin-$1$ analogue to the spin-$1/2$ XXZ-model. We expect to
	find a Kosterlitz-Thouless phase transition at the $SU(3)$ symmetric
	point $\Delta J_3=0$.
\item[(5)]
	The familiar spin-$1$ XXZ-Hamiltonian -- i.e. (5.1) without the
	biquadratic term --
	does not belong to the aforementioned submanifold with degenerate
	groundstate. Of course this does not mean that these Hamiltonians are
	definitely noncritical. We were simply unable to find an operator
	of the Lieb-Schultz-Mattis type for these Hamiltonians.
\item[(6)]
	For $m=4$ --i.e. $s=3/2$-- we found in the $14$ dimensional manifold
	of the anisotropy parameters an $11$ dimensional submanifold, where
	the groundstate is at least twofold degenerate. On a smaller $9$
	dimensional submanifold the groundstate turned out to be at least
	threefold degenerate.
\item[(7)]
	Extension to higher spin cases is straightforward.
\end{itemize}
\section*{Acknowledgments}
It is a pleasure to thank H. Grosse and M. Karbach for discussions.
\begin{appendix}
\appendix
\section{Some properties of $SU(3)$ and $SU(4)$ generators}
The Gell-Mann matrices $\lambda_A, A=1,...,8$:
\begin{eqnarray}
 	\lambda_{1} &=& \left ( \begin{array}{ccc}
		0 & 1 & 0 \\ 1 & 0 & 0 \\ 0 & 0 & 0
            		\end{array} \right ) ,\,
  	\lambda_{2} = \left ( \begin{array}{ccc}
		0 & -i & 0 \\ i & 0 & 0 \\ 0 & 0 & 0
			\end{array} \right ) ,\,
 	\lambda_{3} = \left ( \begin{array}{ccc}
		1 &  0 & 0 \\ 0 & -1 & 0 \\ 0 & 0 & 0
            		\end{array} \right ) ,\, \nonumber  \\  \nonumber \\
 	\lambda_{4} &=& \left ( \begin{array}{ccc}
		0 & 0 & 1 \\ 0 & 0 & 0 \\ 1&  0 & 0
            		\end{array} \right ) ,\,
 	\lambda_{5} = \left ( \begin{array}{ccc}
		0 & 0 & -i \\ 0 & 0 & 0 \\ i&  0 & 0
            		\end{array} \right ) ,\,
 	\lambda_{6} = \left ( \begin{array}{ccc}
		0 & 0 & 0 \\ 0 & 0 & 1 \\ 0&  1 & 0
            		\end{array} \right ) ,\, \\ \nonumber \\
 	\lambda_{7} &=& \left ( \begin{array}{ccc}
		0 & 0 & 0 \\ 0 & 0 & -i \\ 0&  i & 0
            		\end{array} \right ) ,\,
 	\lambda_{8} = \frac{1}{\sqrt{3}}\left ( \begin{array}{ccc}
		1 & 0 & 0 \\ 0 & 1 & 0 \\ 0&  0 & -2
            		\end{array} \right ), \hspace{3cm}  \nonumber
\end{eqnarray}
define the fundamental
(``quark'') representation of the $SU(3)$ Lie algebra:
\begin{equation}
	[\lambda_A,\lambda_B]=2i\sum_{C=1}^8f_{ABC}\lambda_C
\end{equation}
with the totally antisymmetric structure constants $f_{ABC}$.
We can identify the matrices
\begin{equation}
	\lambda_2=s_3,\quad \lambda_5=-s_2,\quad \lambda_7=s_1
\end{equation}
with the $O(3)$ generators $s_l,l=1,2,3$. In contrast to the spin-1
matrices $s_l$, the Gell-Mann matrices close under anticommutation as
well -- like the Pauli matrices:
\begin{equation}
	\{\lambda_A,\lambda_B\}=\frac{4}{3}\delta_{AB}+
	\sum_{C=1}^8 d_{ABC}\lambda_C,
\end{equation}
with the totally symmetric structure constants $d_{ABC}$.

Using the commutation- and anticommutation relations (A.2) and (A.4)
one can express the biquadratic form:
\begin{equation}
	\biglb(\vec{s}(x)\vec{s}(x+1)\bigrb)^2 = \frac{1}{2}
	\sum_{A\neq 2,5,7} \lambda_A(x)\lambda_A(x+1) -
	\frac{1}{2}\sum_{A= 2,5,7} \lambda_A(x)\lambda_A(x+1),
\end{equation}
in terms of bilinears of the Gell-Mann matrices.
The Lieb-Schultz-Mattis construction in Sec. II relies on the following
commutation relations and transformation properties:
\begin{equation}
	[\lambda_A(x),\lambda_8(x)]=0,\quad A=1,2,3,
\end{equation}
\begin{equation}
	[\lambda_4(x)\lambda_4(x+1)\pm \lambda_5(x)\lambda_5(x+1),
	\lambda_8(x)\pm \lambda_8(x+1)]=0,
\end{equation}
\begin{equation}
	[\lambda_6(x)\lambda_6(x+1)\pm \lambda_7(x)\lambda_7(x+1),
	\lambda_8(x)\pm \lambda_8(x+1)]=0,
\end{equation}
\begin{equation}
	U(x)\biglb(\lambda_4(x)\pm i\lambda_5(x)\bigrb) U^+(x) =
	\exp\biglb( \pm i\phi(x) \bigrb)
	\biglb( \lambda_4(x)\pm i\lambda_5(x)\bigrb),
\end{equation}
\begin{equation}
	U(x)\biglb( \lambda_6(x)\pm i\lambda_7(x)\bigrb) U^+(x) =
	\exp\biglb( \pm i\phi(x)\bigrb)
	\biglb( \lambda_6(x)\pm i\lambda_7(x)\bigrb),
\end{equation}
where
\begin{equation}
	U(x)=\exp\left(i\frac{\phi(x)}{\sqrt{3}}\lambda_8(x)\right).
\end{equation}
Finally let us list the $SU(4)$ generators $(\lambda_A)_{jk},A=1,...,15;
j,k=1,2,3,4$ as they were used in Sec. IV. The generators $A=3,8,15$
form the Cartan subalgebra:
\begin{equation}
	\lambda_3=e_{22}-e_{33},
\end{equation}
\begin{equation}
	\lambda_8=\frac{1}{\sqrt{3}}(e_{22}+e_{33}-2e_{44}),
\end{equation}
\begin{equation}
	\lambda_{15}=\frac{1}{\sqrt{6}}(-3e_{11}+e_{22}+e_{33}+e_{44}),
\end{equation}
where $e_{jk}$ are $4\times 4$ matrices with only one nonzero element
in the j'th row and k'th column. The nondiagonal generators can be
expressed in terms of raising and lowering operators $e_{jk}$:
\begin{equation}
	\lambda_A=\frac{1}{2}(e_{jk}+e_{kj}),\quad A=1,4,6,9,11,13,
\end{equation}
\begin{equation}
	\lambda_A=\frac{-i}{2}(e_{jk}-e_{kj}).\quad A=2,5,7,10,12,14.
\end{equation}
The relation between the indices $A$ and $(jk)$ can be read of TABLE I:

\begin{table}
\begin{tabular}{ccccccccccccc}
        A & 1 & 2 & 4 & 5 & 6 & 7 & 9 & 10 & 11 & 12 & 13 & 14 \\
     (jk)&(23)&(23)&(24)&(24)&(34)&(34)&(12)&(12)&(13)&(13)&(14)&(14)
\end{tabular}
\end{table}

\end{appendix}

\end{document}